\documentclass[11pt,twoside]{article}
\usepackage{asp2010}
\usepackage{natbib}

\resetcounters

\begin{document}
\bibliographystyle{asp2010}

\title{CIWS-FW: a Customizable Instrument Workstation Software Framework for instrument-independent data handling}
\author{Vito~Conforti,$^1$ Massimo~Trifoglio,$^1$ Andrea~Bulgarelli,$^1$ Fulvio~Gianotti,$^1$ Enrico~Franceschi,$^1$ Luciano~Nicastro,$^1$ Andrea~Zoli,$^1$ Mauro~Dadina,$^1$ Ricky~Smart,$^4$ Roberto~Morbidelli,$^4$ Marco~Frailis,$^2$ Stefano~Sartor,$^2$ Andrea~Zacchei,$^2$ Marcello~Lodi,$^3$ Roberto~Cirami,$^2$ and Fabio~Pasian$^2$ (the CIWS collaboration)}
\affil{$^1$INAF-IASF Bologna, Via P. Gobetti 101, Bologna, I40129, Italy}
\affil{$^2$INAF-Osservatorio Astronomico di Trieste, Via G.B. Tiepolo 11, Trieste, Italy}
\affil{$^3$Telescopio Nazionale Galileo-INAF-FGG, Rambla José Ana Férnandez Pérez, 7 38712 Breña baja, España }
\affil{$^4$INAF-Osservatorio Astrofisico di Torino,  Strada Osservatorio 20, 10025 Pino Torinese (TO), Italy}

\begin{abstract}
The CIWS-FW  is  aimed at providing a common and standard solution for the storage, processing and quick look at the data acquired from scientific instruments for astrophysics. The target system is the instrument workstation either in the context of the Electrical Ground Support Equipment for space-borne experiments, or in the context of the data acquisition  system for instrumentation. The CIWS-FW core includes software developed by team members for previous experiments and provides new components and tools that improve the  software reusability, configurability and extensibility attributes. The CIWS-FW mainly consists of two packages: the data processing system and the data access system. The former provides the software components and libraries to support the data acquisition, transformation, display and storage in near real time of either a data packet stream and/or a  sequence of data files generated by the instrument. The latter is a meta-data and data management system, providing a reusable solution for the archiving and retrieval of the acquired data. A built-in operator GUI allows to control and configure the IW. In addition, the framework provides mechanisms for system error and logging handling. A web portal provides the access to the CIWS-FW documentation, software repository and bug tracking tools for CIWS-FW developers. 
We will describe the CIWS-FW architecture and summarize the project status. 
\end{abstract}

\section{Introduction}
Many data acquisition and archiving software have been developed in the recent years for the instrument workstation (IW) to be procured by the instrument team both in the AIV/AIT (Assembly Integration Verification and Test) and the commissioning and operation contexts for ground-based telescope and space born experiments in astrophysics. 
The aim of the project is to promote the IW software reuse providing a framework that allows developers, involved in astrophysics project/mission, to easily implement the IW software. It has been conceived to improve configurability and extensibility of the software developed by the team in the past years, e.g. in AGILE\citep{agile}, Planck\citep{planck}, TNG\citep{tng}, GSCII\citep{gsc2}.  
To introduce newbies developers, the CIWS-FW is presented on a web site (http://ciws-fw.iasfbo.inaf.it/ciws-fw) organized in the following sections: (i) CIWS-FW learning: tutorials, handbooks, examples, programmer's guide; (ii) CIWS-FW download: software repository, installation guide;  (iii) CIWS-FW community: software bug tracking and forum.

\section {CIWS-FW concept}
\begin{figure}[!ht]
	\centering
 	\includegraphics[width=0.7\textwidth]{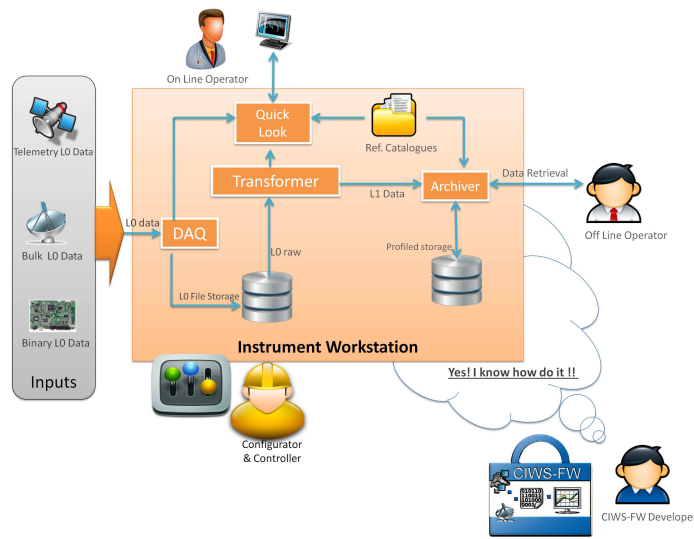}
	\caption{CIWS-FW Model View}
	\label{fig:concept}
\end{figure}

The CIWS-FW concept is depicted in \figurename~\ref{fig:concept}.

The IW software to be developed using the CIWS-FW should be configurable for near real time acquisition of packet data streams and file data streams that should be transformed and archived according to the data models defined by the user. The IW provides configuration and controller GUI to control and monitor the operations, an on-line operator GUI to display in near real time the acquired data, an off-line operator GUI to retrieve and display archived data. The CIWS-FW software aids the IW developer with basic software components and tools that can be extended in order (i) to model the data, (ii) to acquire and store the source data (L0) in raw format, (iii) to generate the transformed data (L1) according to the user data model, (iv) to archive and retrieve them according to the meta-data specified by the user, (iv) to implement the operators GUI. The configuration and controller GUI is included in the framework as a built-in component ready to configure and control the IW software.

\section {CIWS-FW Architecture}
\begin{figure}[!ht]
	\centering
 	\includegraphics[width=0.7\textwidth]{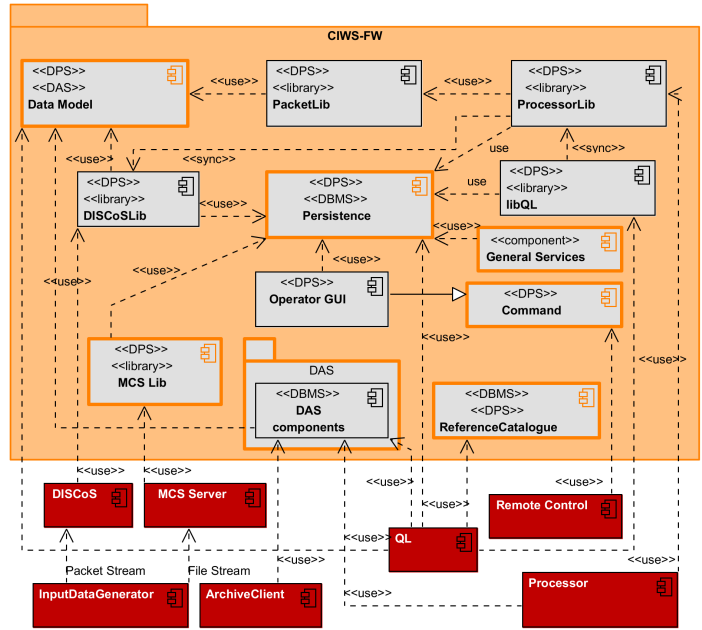}
  \caption{CIWS-FW Architecture}
	\label{fig:architecture}
\end{figure}

\figurename~\ref{fig:architecture} reports the CIWS-FW components diagram.

This architecture represents an upgrade of the IW software developed by the team for recent space missions.  The highlighted gray components have been added in order to improve the configurability and extensibility attributes of the system and to offer new capabilities. The gray components are part of the CIWS-FW package which includes the data processing system (DPS) and the DAS  whereas the red components are the instrument specific components that the developer who is going to use CIWS-FW must implement. The Data Model component provides the DDLs grammar as XML schema definition (XSD) and are exploited by the PackeLib \citep{packetLib} to decode L0 data, by the libQL component for quick look purposes, and by the DAS components  to handle the L1 data. The L0 DDL grammar should be compliant with the XML telemetric and command exchange (XTCE)\citep{xtce} standard, hence it should be able to support the European Space Agency packet utilization standard (ESA PUS)\citep{esa}. Details on the L1 DDL grammar, and on the DAS package in general, are given in \citep{das}. The developer is in charge of providing the related XML files validated on the DDLs data model. The input data are received by the DISCoS\citep{discos} component (which uses the DISCoSLib component) in the case of a packet stream,  and by the MCS Server\citep{mcs} component in the case of file streams. The Command component handles the commands that change the state of the IW system. The commands could be run either through  the Operator GUI Component or the Remote Control component. The Operator GUI allows the Developer to set/get the system configuration, and the Operator to configure the data taking context. In the packet stream case, the developer implements one Processor component for each input data type. Each Processor component is activated by a DISCoS component upon receiving a new data packet of the incoming stream. Through the ProcessorLib and PacketLib components, each Processor transforms the L0 data into the L1 data. The transformation is performed according to the L1 DDL model defined by the developer. The cfitsio  library  is used to archive the L1 data into FITS files. Specific API are provided by the DAS component to archive the L1 profiled data (meta-data are also associated with the file in a database). The DAS API can further be used to implement data access and retrieval on the meta-data for the ArchiveClient component. In the file stream case it is assumed a command-oriented connection which is served by MCS Server. The MCSLib component provides to the developer the mechanisms to define the action to be performed for each command. The quick look (QL) component is able to monitor, annotate, and display the L0 and L1 data,  either in near real time or off-line. The ReferenceCatalogue component allows the QL component to access custom astronomical source catalogues in order to accomplish tasks like producing a simulated field or cross-matching detected sources. The list of catalogues the user has access to is very large and they can either reside on the local machine or can be accessed from the internet. We use the HTM pixelization scheme to index MySQL tables containing the catalogues data \citep{catalogues}. Additional database tables store metadata information used by SQL functions and procedures as well as by a C++ API library. This structure makes the CIWS source catalogues self-consistent and easily expandable. 
The data acquisition synchronization mechanisms, system configuration, and reporting are provided by the Persistence component, which makes use of the CIWS-FW system DB. General Services provided by the framework includes for system error and logging handling mechanism.

\section{Conclusions}
The CIWS adopted an iterative and incremental software life cycle model. At the time of writing we are in the production phase. We have completed the requirements definition, and almost  completed the design. We have implemented and deployed prototypes of the Data Model, DAS, Persistence, Operator GUI,  and Reference Catalogues. A first stable release is expected in the upcoming months. The CIWS-FW web site is on line at http://ciws-fw.iasfbo.inaf.it/ciws-fw.  The software repository and bug tracking system is available at  http://redmine.iasfbo.inaf.it/projects/ciws-fw. The CIWS-FW is being developed through the TECNO-INAF-2010 technological project funded by INAF.

\bibliography{P065}

\end{document}